\documentclass[preprint,prb,preprintnumbers]{revtex4}
\usepackage[latin9]{inputenc}
\usepackage{dcolumn}
\usepackage{float}
\usepackage{natbib} 
\usepackage{graphicx}

\begin{document}

\title{Homogeneous bubble nucleation in water at negative pressure: A Voronoi polyhedra analysis}

\author{Jose L. F. Abascal, Miguel A. Gonzalez, Juan L. Aragones and C. Valeriani}

\affiliation{Departamento de Qu\'{\i}mica F\'{\i}sica, Facultad de Ciencias 
Qu\'{\i}micas, Universidad Complutense de Madrid, 28040 Madrid, Spain}

\date{\today}

\begin{abstract}
We investigate  vapor bubble nucleation in metastable TIP4P/2005 water at
negative pressure via the Mean First Passage Time (MFPT) method using the
volume of the largest bubble as a local order parameter. We identify the
bubbles in the system by means of a Voronoi-based analysis of the Molecular
Dynamics trajectories. 
By comparing the features of the tessellation of liquid water at ambient
conditions to those of the same system with an empty cavity we are able to
discriminate   vapor (or interfacial) molecules from the bulk ones.
This information is used to follow the time evolution of the largest bubble 
 until the system cavitates  at 280 K above and below the spinodal line.
At the pressure above the spinodal line, the MFPT curve shows the expected
shape for a moderately metastable liquid from which we estimate 
the bubble nucleation rate and the size of the critical cluster.
The nucleation rate estimated  using Classical Nucleation Theory 
turns out to be about 8 order of magnitude lower than the one 
we compute by means of MFPT. 
The behavior at the pressure below the spinodal line, where the liquid is 
thermodynamically unstable, is 
remarkably different, the MFPT curve being a monotonous function without any  
inflection point. 
\end{abstract}

\maketitle

\section{Introduction}

Bubble nucleation is a widespread phenomenon in our daily life,  relevant to  processes such as 
explosive boiling\cite{shusser99} and  sonochemistry\cite{suslick90}.
Despite its technological relevance, the mechanism of nucleation 
of vapor bubbles from homogeneous metastable liquids is still not entirely understood.
Classical Nucleation Theory (CNT) is usually employed to predict the nucleation
rate, both in bubble nucleation and in  vapor condensation experiments. 
Early  measurements of  vapor condensation  estimated nucleation rates 
consistent with the CNT predictions.\cite{blander75}
Regarding   bubble nucleation, measurements on different 
materials\cite{viisanen94,wagner84,rudek99,lihavainen01,kim04,gharibeh05,herbert06,caupin05}
have shown  that, even though Classical Nucleation Theory correctly  predicts the temperature dependence 
of the bubble nucleation rate, it yields   nucleation rates
that are too low at the lowest temperatures. 
Delale et al.,\cite{delale03} using a phenomenological estimate of the minimum
work of bubble formation, computed 
 the steady-state bubble nucleation rate and found that, as with the CNT
estimates, there was a large discrepancy between the           computed and measured
bubble-nucleation rates.
Recently, El Mekki Azouzi et al.\cite{azouzi2012}, performing experiments of liquid water  under large mechanical tension,
used the cavitation statistics to get the free-energy barrier. 
The authors observed that both the free energy and the volume of the critical bubble were well described by CNT
whereas the surface tension was reduced. 

From a theoretical and numerical point of view, Zeng and Oxtoby\cite{zeng91}
used Density Functional Theory to estimate the bubble nucleation rate in a
super-heated Lennard Jones system and concluded that CNT underestimated the
nucleation rate by more than 15 orders of magnitude.
One reason for it was that CNT neglects curvature corrections to the surface
free-energy of the bubble\cite{talanquer95}.
In 1999, Shen and Debenedetti\cite{shen99} performed a  computer simulation
study of bubble nucleation from a metastable super-heated Lennard Jones
system.
Using the  vapor density as a global order parameter, they        found that
the critical nucleus was a large system-spanning cavity, in contrast to the
spherical cluster predicted with CNT.
Later on, Wang et al\cite{wang09} reported a Molecular Dynamic study 
of homogeneous bubble nucleation for the same  Lennard-Jones fluid.  
Using Forward Flux Sampling and the volume of the largest bubble as a local
order parameter, they        computed the bubble nucleation rate at a given
super-heating and obtained 10$^{-22} \sigma^{-3} \tau^{-1}$. 
At the same conditions, the CNT estimate could vary from 10$^{-22} \sigma^{-3} \tau^{-1}$ 
to  10$^{-36} \sigma^{-3} \tau^{-1}$, depending on the value of the interfacial 
tension (whether obtained  from MD or DFT calculations) used to estimate the free-energy barrier height.
The authors also observed that local temperature fluctuations correlated strongly with bubble formation 
(mechanism not taken into account in CNT) and that, 
contrary to Ref.~\onlinecite{shen99}, cavitation started from compact bubbles. 
Recently, Meadley and Escobedo\cite{meadley12}
studied the same Lennard Jones metastable fluid, undergoing not only super-heating 
but also over-stretching (negative pressure). 
The authors used several rare-event  numerical techniques and both a local (the
volume of the largest bubble) and a global order parameter (the global vapor density) to follow the
liquid to vapor phase transition and estimated the nucleation rate and the
free-energy barrier.
They found that the free-energy barrier was higher when projected over the
bubble volume rather than over the vapor density  and concluded that the 
former  was a more ideal reaction
coordinate. When analyzing the shape of the growing vapor bubble, they observed
bubbles with non-spherical shapes and irregular and undulating surfaces. 

In contrast to the work done to unravel the bubble nucleation mechanism in
simple fluids (such as Lennard Jones), little effort has been made so far  
to understand the mechanism behind the liquid to vapor transition in molecular liquids such as water.
The goal of the present manuscript is to study bubble nucleation in numerical
simulations of TIP4P/2005 water at negative pressures using the size of the 
largest bubble as a local order parameter.
The bubble volume is obtained by means of a Voronoi-based algorithm.
Knowing the location of the spinodal line for TIP4P/2005 water at negative
pressures,\cite{master_mac} we study bubble nucleation for two state points
located, respectively, slightly above and below the spinodal.
In both cases the system cavitated spontaneously.
Therefore, the {\it Mean First Passage Time} (MFPT) 
method\cite{wedekind07,chkonia09} seemed to be adequate to compute the bubble
nucleation rate and to estimate the volume of the critical bubble.
MFPT has been successfully applied to study liquid condensation from 
metastable vapor in a Lennard-Jones system\cite{wedekind09,wedekind07c} and 
from metastable TIP4P/2005 water\cite{perez11}. 
The application of the MFPT technique to bubble nucleation ultimately relies in
the investigation of the time evolution of the bubble growth.
The detection of bubbles in computer generated configurations of a metastable
liquid is more complex than it may appear at first.
A number of procedures have been proposed to distinguish liquid-like 
from  vapor-like particles.\cite{senger99,wedekind07b,vanmeel12}
A bubble is a region of  vapor within the liquid and since the  vapor
is much less dense than the liquid (except in the vicinity of the critical
point), in principle, one might think of a bubble as a ``void" region.
Given the small size of the bubble, it is difficult to know to what extent the
interface affects the structure of the  vapor molecules.
The problem is more complex for a network forming liquid such as water
whose structure is not determined exclusively by packing effects since
the hydrogen bond network is a structure with many voids.
Because of this, the fixed cutoff criterion of Ref.\onlinecite{tenwolde98}
which has been successfully employed in the analysis of the condensation of a
metastable Lennard-Jones vapor might not work in the case of water.
We thus face a challenging geometrical problem.

The procedure to tag the molecules as either liquid or  vapor may be
performed in three steps: {\em 1)} A partition of the space between the molecules into regions, {\em
2)} An assignment of each region as belonging to a "vapor" (or interfacial) or a
liquid molecule, and {\em 3)} A clustering of the  vapor molecules into bubbles.
To solve the first step, the Voronoi (also known as Dirichlet) tessellation
is a natural choice leading to an unambiguous partition of the system.
The Voronoi construction is a partition of the space based on the distance to a
set of points.
Each region contains exactly one generating point and every point in a
given region is closer to its generating point than to any other.
The Voronoi tessellation has been found useful in many scientific and technical
applications.\cite{okabe00,voroeppstein,vorowiki,karch03}
When the space is Euclidean and the set of points are physical or chemical
particles, it allows to univocally assign a given region to every particle ---a
polygon in 2D and a polyhedron in 3D--- thus providing structural information
of the system.
To our knowledge, the first application of the tessellation to elucidate the
structure of a chemical-physics system was done by Bernal in 1964.%
\cite{bernal64}
Since then a number of studies have used the Voronoi tessellation to investigate
the structure of fluids, glasses and solids.\cite{cape81,gilmontoro93,%
gilmontoro94,yang02, gedeon02,jedlovszky04,xu07,canales09,gedeon11,skvor11,%
idrissi11,swart11}
Other typical examples of the utility of the Voronoi construction are the
analysis of solvation shells,\cite{david82,neumayr10,voloshin11,haberler11}
clustering and chemical association.%
\cite{gilmontoro94,jedlovszky00,gomezalvarez11,hoedl11}
The Voronoi construction has also been employed to define the interface between
two systems\cite{usabiaga09,bouvier09,isele-holder12} and to obtain structural
information of pure water.\cite{geiger92,shih94,jhon06,malenkov07,yeh99,
jedlovszky00,jedlovszky08,stirneman12} and water solutions\cite{gilmontoro94,va%
isman94,jedlovszky99,zapalowski00,zapalowski01,idrissi08,neumayr09,schroeder09}
More closely related with the goals of this work is its application to
detect different types of cavities such as pores, pockets, clefts, channels and
docking sites.\cite{david86,david88,lewis89,jedlovszky00,alinchenko05,kim06b,%
petrek07,kim08,sonavane08,chakraborty11,wang11}
In fact, Fern {\em et al.}\cite{fern07} have already used the Voronoi
tessellation as a tool for distinguishing whether a particle belongs to a
condensed or  vapor phase in two-phase simulations of ethanol.

The Voronoi tessellation can be performed trivially in crystalline systems
but it becomes increasingly complex in disordered systems.%
\cite{gilmontoro93,gilmontoro94,kim05,kim06,sonavane08}
In fact, when the system is strongly heterogeneous (as is the case of bubbles
within a metastable liquid) many algorithms fail.
In this work we use an algorithm especially designed to deal with inhomogeneous
fluids     which has been successfully applied to interpret the structure of
quenched liquids\cite{gilmontoro93} and electrolyte
solutions.\cite{gilmontoro94}
Once the tessellation is done, the properties of the Voronoi polyhedra
(VP) associated to each particle can be used to distinguish between liquid and
vapor molecules.
The choice of the parameters that denote a molecule as belonging to the  vapor
phase is somewhat arbitrary.
Therefore, as a test case, we have investigated the properties of a system
consisting of liquid water with an artificially created spherical cavity.
Once the  vapor molecules are identified with the Voronoi construction, it is a
trivial task to cluster them into one or more bubbles because the Voronoi
tessellation directly provides the list of neighbors of each molecule. 
We then use the volume of the largest bubble as the order parameter and apply
the MFPT technique to study the spontaneous bubble nucleation.\cite{wedekind07}
The analysis of the evolution of the volume of the largest bubble provides the
desired relevant features of the nucleation, in particular the nucleation rate
and the size (volume) of the critical bubble.
In this work we will show that the overall procedure is quite robust and 
yields a consistent description of the bubble nucleation.

The main drawback of the overall method is that the Voronoi tessellation is far
from trivial for highly inhomogeneous systems. We have thus dedicated a section
to describe the algorithm used for the tessellation and another one to
investigate what are the relevant parameters of the tessellation that enable us
to track down the bubble growth.
The paper is organized as follows. Section II describes some methodological
issues, in particular, the algorithm employed for the Voronoi construction
and a brief description of the simulations. Section III analyzes which
are the relevant parameters of the Voronoi Polyhedra that allow to detect a bubble within
metastable liquid water. Finally, section IV reports the application of the
MFPT technique to two state points, one above and the other below the
liquid- vapor spinodal line of TIP4P/2005 water.

\section{Methods}

\subsection{Algorithm used for the Voronoi tessellation}

The Voronoi tessellation involves the calculation of the intersection
of the planes normal to the line joining a given particle with the 
remaining ones. Usually, 
in simple liquids, the number of Voronoi neighbors is less than twenty.
It is then a waste of resources evaluating the normal planes for all 
particles in the system.
Thus, it is customary to perform the calculations for a reduced set of
particles, the so called "candidate neighbors".

\begin{center}
\begin{figure}[!ht]
\caption{Particles A and B have not been identified by the grid as Voronoi
neighbours. As a consequence the filled region would be ascribed to both
particles (and it would be counted twice).
Since A and B have C and D as common neighbors, the expansion of the
initial list for A and B to include particles with a given number of common
neighbors (see step 2 in the text) would allow to divide the filled volume
among A and B.}
\includegraphics*[clip,scale=0.6]{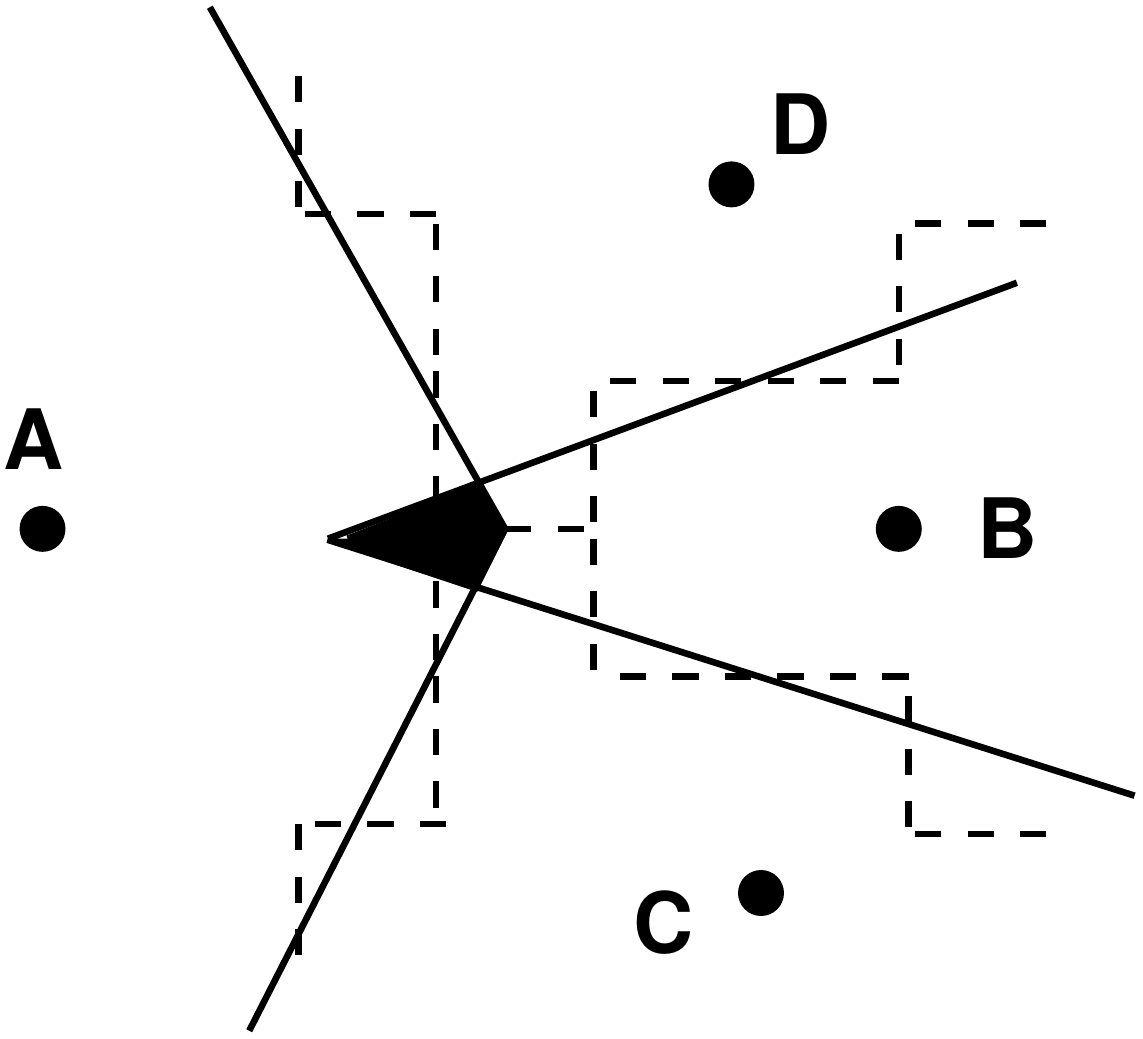}
\label{fig:voro_grid}
\end{figure}
\end{center}

The list of candidate neighbors of each particle is usually obtained using a
cutoff radius.
This procedure works well for common liquids but fails for inhomogeneous
systems\cite{gilmontoro93,gilmontoro94} because some actual Voronoi neighbors
are relatively distant from the reference particle.
In the case of a bubble some Voronoi neighbors may be found at distances
of the order of the bubble diameter so the number of candidate neighbors may well
exceed a hundred particles.
Thus, it is important to devise a different way to obtain a reliable
list of candidate neighbors without the computational cost associated to
a very large cutoff.
As in Ref.~\onlinecite{gilmontoro93}, the procedure we follow can be described in 
few steps.\\
{\bf Step1:} First, we obtain  a first approximation of each particle's list of Voronoi 
candidate neighbors. This is done by dividing the system in boxes with the use of a grid,
each box being assigned to its closest particle. Therefore, two particles are nearest 
candidate neighbors if at least two boxes (each box belonging to a  particle) are adjacent.
Rigorously, this stage of the algorithm only works in the limit of zero
thickness grid. Thus, we need to improve this initial list of candidates. \\
{\bf Step 2:} To amend the calculated set of candidate neighbors, 
the atoms sharing a certain number of common neighbors with a given one are
recursively added to its list of candidates (see Fig.~\ref{fig:voro_grid}).
At the end of this stage, we have generated a set of candidates for
each particle that hopefully includes all the actual Voronoi neighbors.\\
{\bf Step 3:}  Having the list of candidate neighbors for all the particles,
we proceed with the tessellation using any standard 
algorithm.\cite{brostow78,finney79,tanemura83,medvedev86}
For this final third step we choose to use a variation of the Finney's 
algorithm\cite{finney79} in order to increase its efficiency (see 
Ref.~\onlinecite{gilmontoro93} for details and a forthcoming 
paper\cite{voro_code}).

The main advantage of the algorithm is its robustness.
However, the large inhomogeneities of our systems have forced us to use it 
choosing parameters in a conservative way.
We have used a 200x200x200 grid and performed the neighbor expansion with a reduced
number (6) of common neighbors. In these conditions, the number of actual 
Voronoi neighbors  can be sometimes close to 40.
With these parameters, we have been able to analyze all the generated 
configurations (about 8$\cdot 10^5$ in total) before the system cavitates.

We have carried out 200 independent Molecular Dynamics simulation runs for each of
the state points of liquid water at 280~K. 
One of them, at p=-2250 bar, is above the spinodal line and the other one, at
p=-2630~bar, is below it (recently, a first approach to the spinodal line 
at low temperatures has been reported for this model\cite{master_mac}).
All simulations have been performed for 500 water
molecules in the isothermal-isobaric $NpT$ ensemble using the Molecular
Dynamics package GROMACS\cite{hess08} with a 1~fs timestep.
Long range electrostatic interactions have been evaluated with the smooth
Particle Mesh Ewald method.\cite{essmann95}
The geometry of the water molecules has been enforced using
{\em ad hoc }constraints.\cite{ryckaert77}
Temperature has been set to the desired value with a velocity rescaling
thermostat\cite{bussi07} and, to keep the pressure constant,
an isotropic Parrinello-Rahman barostat has been applied.\cite{parrinello81}
We have used the TIP4P/2005 model\cite{abascal05b} to describe the interactions
between water molecules.
This model provides a quantitative account of many water
properties,\cite{vega09,vega11} including not only the well known thermodynamic
anomalies,\cite{pi09} but also the dynamical ones.\cite{gonzalez10}
Of particular relevance for the purposes of this work is the high quality
of the predictions for the  vapor-liquid equilibria.\cite{vega06}
It is finally to be mentioned that the TIP4P/2005 model has been successfully
used to study the nucleation of liquid droplets within a metastable
vapor.\cite{perez11}

\section{Voronoi tessellation as a tool for detecting bubbles}

\subsection{Test case: Liquid water with an empty cavity}
Our goal is  to detect the formation and growth
of one or more bubbles within metastable liquid water.
Therefore, we need a criterion to distinguish between  vapor and liquid
particles.  Given the density difference between  vapor and liquid, a bubble
will contain few  vapor molecules surrounded by liquid ones.
Thus, to a first approximation, a bubble appears as a ``void'' in the liquid.
It is then probably simpler to detect the bubble using the location of the interfacial 
molecules.
To better characterize the interfacial molecules, we analyst the differences
between the Voronoi Polyhedra (VP) obtained for a given number of
configurations of bulk water at 298~K and 1~bar and the VP computed for  the
same configurations where we have artificially created an empty spherical
cavity by removing 30 water molecules (this approximately corresponds to a
0.6~nm cavity radius and a 0.90-0.95nm$^3$ volume). 
Using the Voronoi tessellation, we compute properties such as the VP volume 
and the non-sphericity parameter (i.e., a measure of the deviation of the Voronoi Polyhedra 
from a spherical shape). 
The differences between these properties have to be attributed to the
interfacial molecules  around the cavity since the remaining particles have
exactly the same environment.
We may then find what are the VP properties that allow to discriminate
interfacial particles (only present in the system with a cavity) from the
bulk ones (common to both systems).
We note that the Voronoi tessellations have been done using only the
positions of the oxygen atoms.

\begin{center}
\begin{figure}[!ht]
\caption{Distribution of VP volumes for TIP4P/2005 water at 298~K, 1~bar.
The difference between the systems is that in one of them an empty 
spherical cavity of 30 molecules has been created.}
\includegraphics*[clip,scale=0.8]{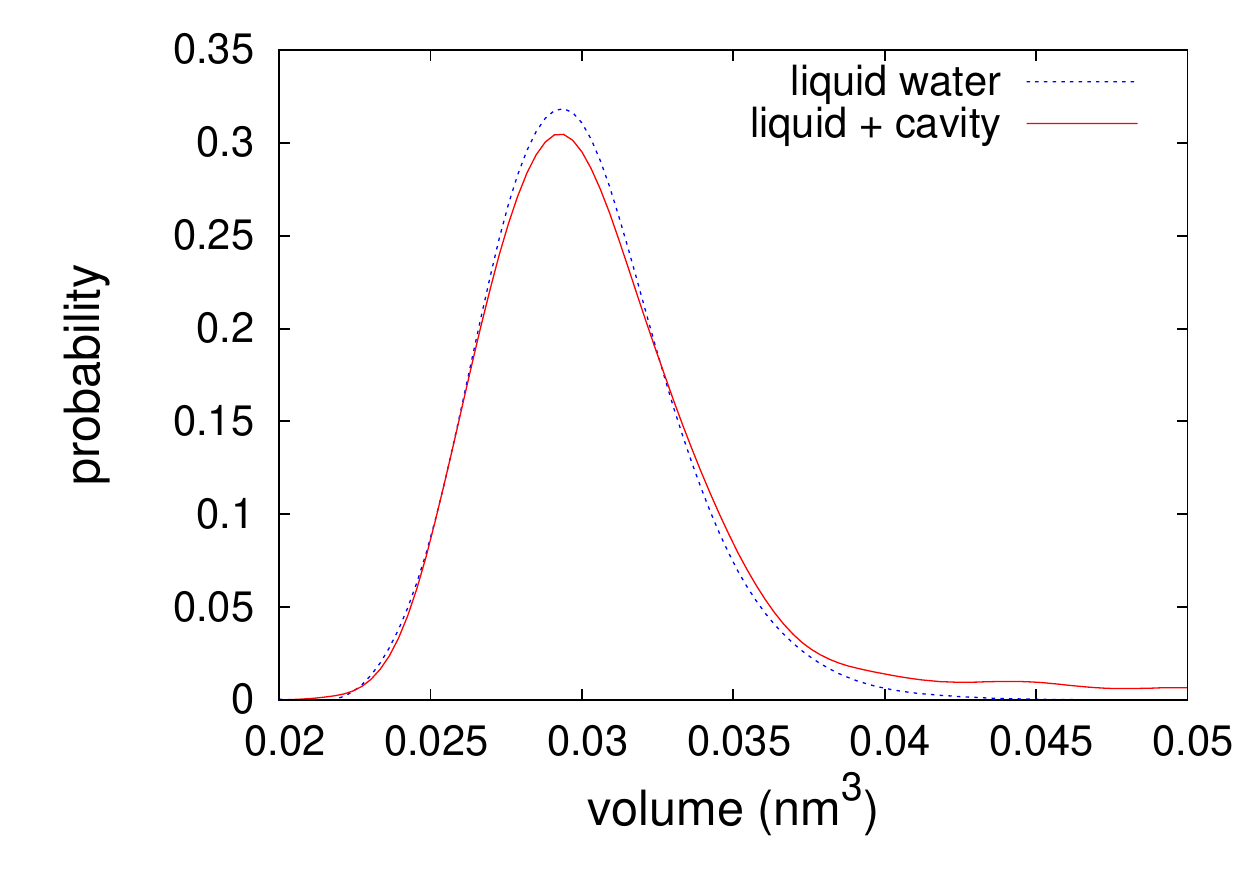}
\label{fig:vol_liq-cavity}
\end{figure}
\end{center}

Figure \ref{fig:vol_liq-cavity} compares the normalized distribution of the VP
volumes for a 500 molecules system (470 in the system with the cavity).
Clearly, the volume distributions are identical (taken into account the
different number of particles in each system) up to a VP volume of about
0.032~nm$^3$.
Larger volumes in the system with the cavity outnumber the corresponding ones
in the bulk due to the fact that the tessellation assigns  the empty space to the
``interfacial" particles.
This increase is progressive but, for volumes between 0.032~nm$^3$ and 
0.04~nm$^3$, it is difficult to label the molecules as either bulk or
interfacial based only on this criterion since these volumes appear in more or
less similar proportions in both systems.
Molecules with VP volumes larger than 0.04~nm$^3$ are almost 
inexistent in the bulk system while they represent a small but significant
part of the total molecules in the system with a cavity.
In order to better distinguish bulk from interfacial molecules, it seems
necessary to use another parameter obtained from the VP tessellation. 
The non-sphericity (or asphericity) parameter has been widely used for that purpose.%
\cite{shih94,yeh99,liao02,karch03,gedeon05,idrissi10,jedlovszky00,stirneman12}
Sometimes the definition of the Voronoi Polyhedra anisotropy is made in terms of the
VP surface and volume but a more rigorous definition should also involve the
mean curvature radius.
Thus, we define the nonsphericity parameter, or anisotropic factor $\alpha$,
with the usual expression valid for convex bodies, namely 
\begin{equation}
\label{eq:self}
   \alpha=RS/3V
\end{equation}
where $S$ and $V$ are the surface and volume of the convex body, respectively,
and $R$ the mean curvature radius.
In some cases, the geometry of the convex body makes difficult the evaluation
of $R$.
This is not the case of a convex polyhedron for which a simple expression
allows the calculation of $R$:
\begin{equation}
    R=\frac{1}{8 \pi} \Sigma l_i \phi_i \: ,
 \end{equation}
where the sum extends over polyhedron edges of length $l_i$ and
$\phi_i$ is the angle between the normal vectors to the intersecting faces.
These magnitudes are trivially evaluated in the VP tessellation, so a rigorous
calculation of $\alpha$ is numerically possible.
As a reference, the nonsphericity parameter of a sphere is 1 and the
distribution of $\alpha$ for a representative Lennard-Jones liquid has a
maximum at about 1.3.\cite{gilmontoro93}
\begin{center}
\begin{figure}[!ht]
\caption{Distribution of VP nonsphericity factors for the systems of Fig.~1}
\includegraphics*[clip,scale=0.8]{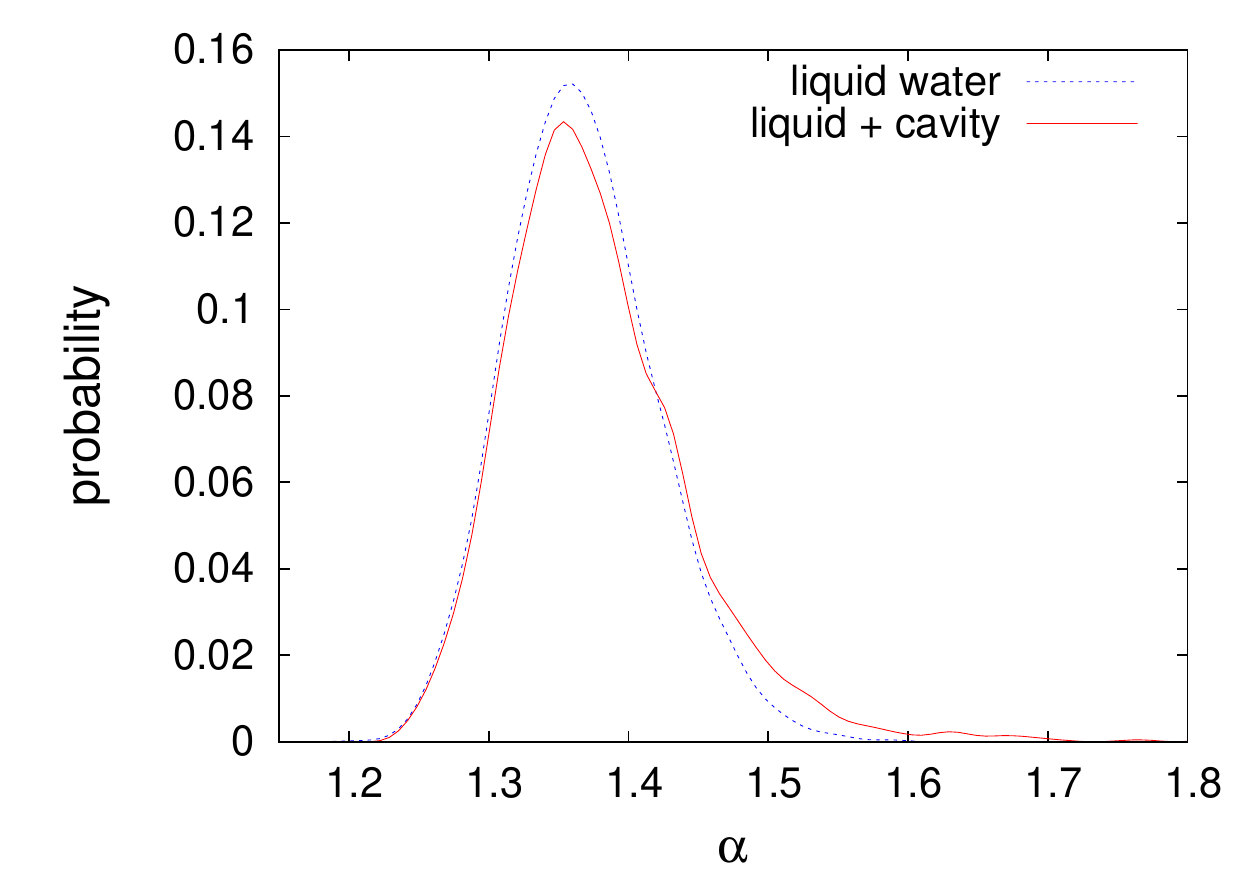}
\label{fig:sph_liq-cavity}
\end{figure}
\end{center}
The nonsphericity factors of the VP for the liquid and the liquid with a cavity
systems are shown in Figure~\ref{fig:sph_liq-cavity}.
The larger anisotropy of the interfacial molecules around the cavity gives rise
to a hump in the distribution function.
But, similarly to the case of the volume distribution, there is a range of 
anisotropies (approximately $1.4<\alpha<1.5$) which may correspond to
either bulk or interfacial molecules making difficult to distinguish
between them.
Only when $\alpha>1.5$ one may be reasonably confident that the nonsphericity
parameter represents interfacial molecules.

\begin{center}
\begin{figure}[!ht]
\caption{Distribution of VP number of faces for the systems of Fig.~1}
\includegraphics*[clip,scale=0.8]{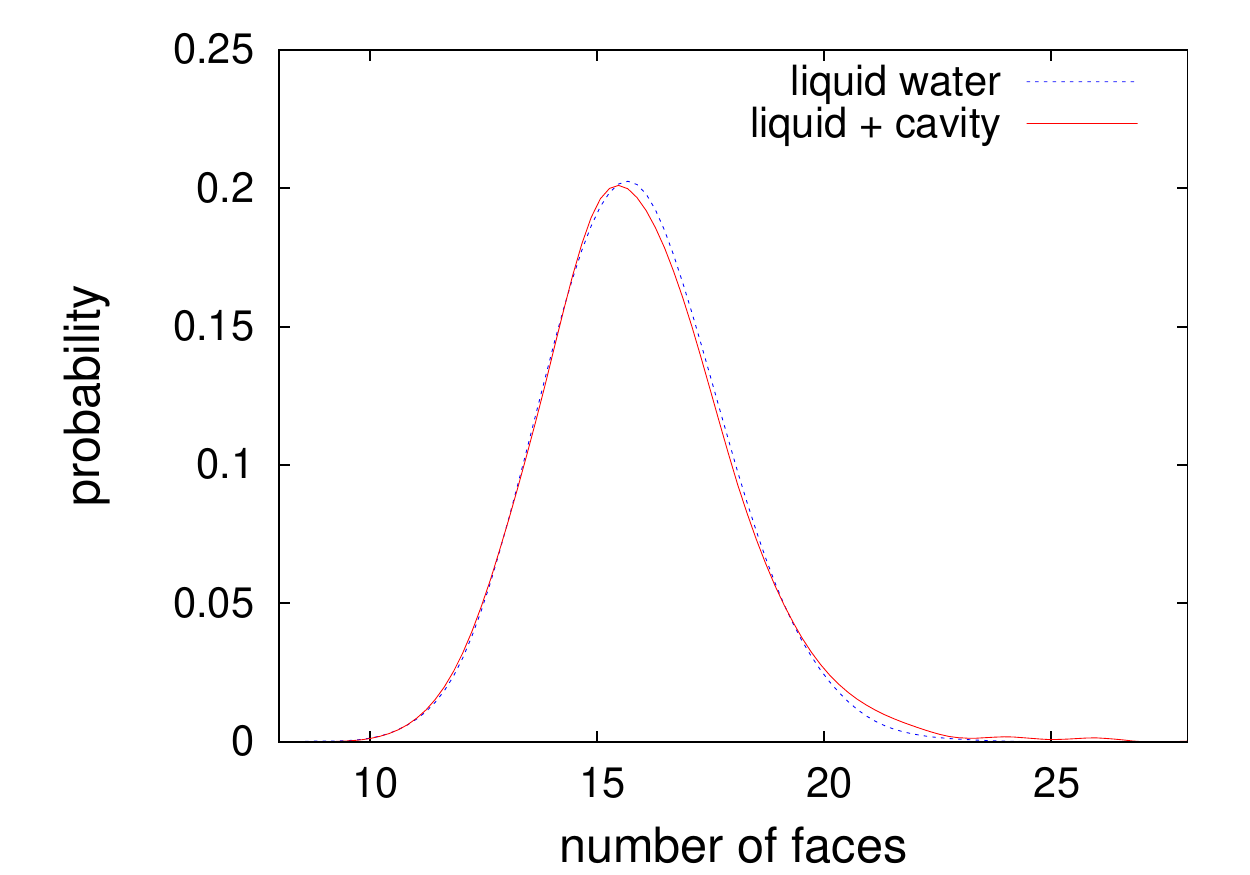}
\label{fig:faces_liq-cavity}
\end{figure}
\end{center}

Therefore, both the VP volume and anisotropy allow to unambiguously
characterize only a part of the interfacial molecules, {\em i.e.,} those with
$V>0.40$~nm$^3$ or $\alpha>1.5$.
Thus, we  need additional parameters in order to improve the particles' assignment 
(whether liquid or  vapor-like).
The VP number of faces has been sometimes proposed as a relevant structural
parameter.
However this seems not to be the case for water.
Figure~\ref{fig:faces_liq-cavity} shows that the distribution of faces is
quite similar in both systems and does not provide any additional information.
It is worth comment that, in liquid water, a non-negligible number
of Voronoi Polyhedra has more than 20 faces and this number is over 30 for some
interfacial molecules, which explains the difficulties in designing a robust
algorithm for the Voronoi tessellation.

\begin{center}
\begin{figure}[!ht]
\caption{Anisotropic factor $\alpha$ as a function of volume for the systems of
Fig.~1. The black line $\alpha=1.5-19*(V-0.04)$ separates bulk molecules from
interfacial molecules.}
\includegraphics*[clip,scale=0.8]{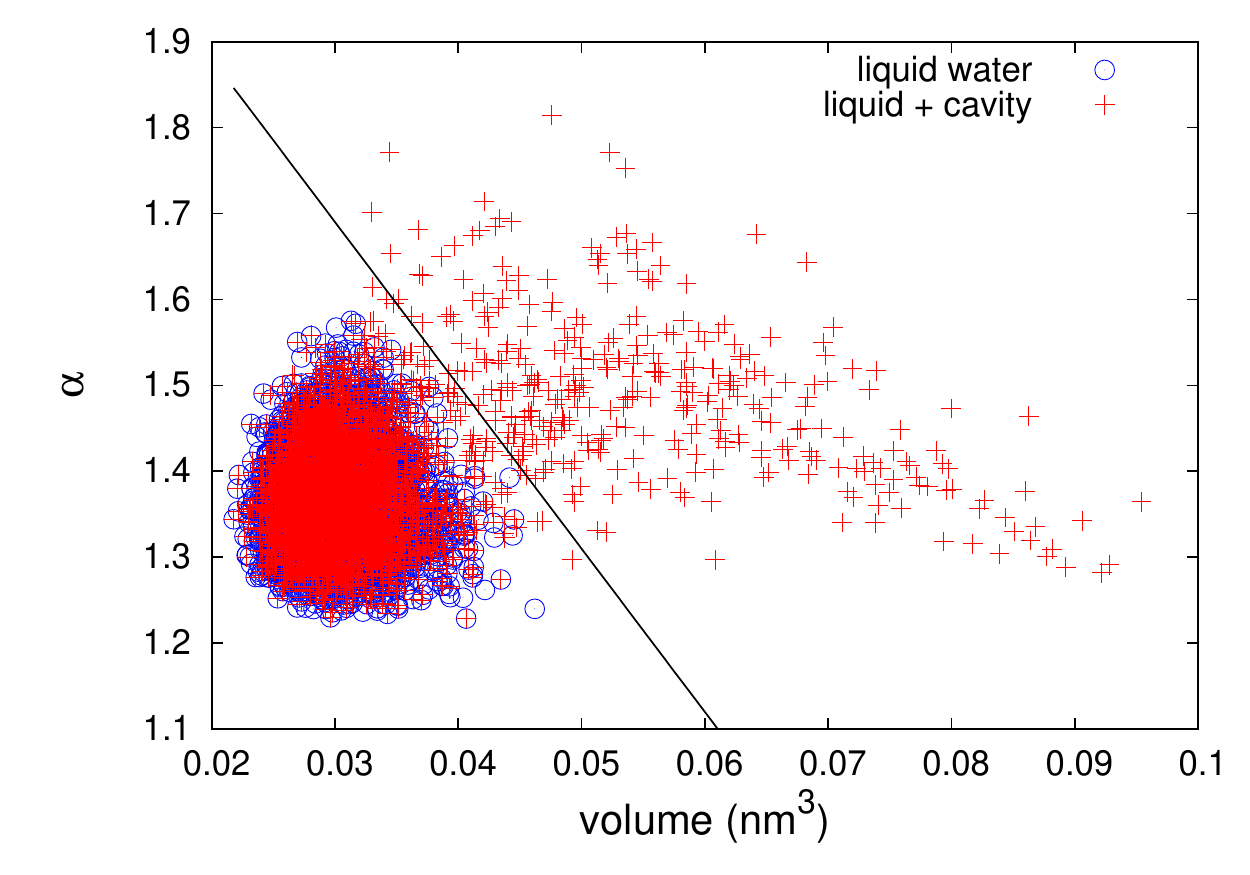}
\label{fig:vep_liq-cavity}
\end{figure}
\end{center}

Since the use of a single parameter, either $V$ or $\alpha$, does not allow an
unequivocal assignment of the interfacial molecules, we suggest to use a
correlation of both parameters in order to enhance the ability to detect them.
Figure~\ref{fig:vep_liq-cavity} represents the values of $\alpha$ for 11
randomly selected configurations plotted against their corresponding volumes.
Notice that the points for the liquid system (without cavity) are grouped in a
 relatively small ``pear-shaped" region of the $\alpha$-$V$ plane.
Most of these points overlap with those of the system with a
cavity because they correspond to the common (bulk) molecules.
On the right side of the figure appear only points coming from the system
with a cavity. 
These must be ascribed to the molecules surrounding the cavity (interfacial
molecules) because their environments are different in both systems.
Therefore, the parameters useful to identify the interfacial molecules are
clearly visible in this plot.
In the figure we have also traced a straight line ($\alpha=1.5-19*(V-0.04)$)
separating two regions: all the liquid-like (bulk) particles lie to the left of
the line while most of the points specific of the system
with a cavity (interfacial molecules) are on the right side of the line.
One could argue that the position of the line could be shifted slightly to the
left but we opted for a conservative approach.
In fact, the line pivots around the point ($v=0.04$~nm$^3$, $\alpha=1.5$) which
was previously shown to clearly indicate that a particle is interfacial.
We also stress that the sample used for the figure is relatively small
(only 11 configurations out of about 500 molecules for each system).
Plotting a larger number of  configurations would imply the appearance of a larger
number of molecules with parameters in the right side of the distribution tails
(with large values of both volumes and anisotropies), and some among them  in the interfacial region.
In summary, we recommend a linear combination of the VP volume and anisotropy $\alpha$ to 
ensure that no bulk particles are tagged as interfacial even if it slightly
underestimates the number of the latter. As shown, 
the use of this parameter is highly discriminating and allows to safely detect
most of the interfacial molecules.

The advantage of the analysis of this system is that it allows us  to compare the
predictions of the Voronoi tessellation with the actual data for the cavity.
The calculated average volume of the cavity is 1.70~nm$^3$, sensibly higher than
the cavity volume which is 0.93~nm$^3$.
But the volume of the bubble, as computed with the Voronoi tessellation, also
includes the volume occupied by  the interfacial molecules which was not taken
into account for the cavity's volume.
Therefore we must subtract the interfacial volume, that consists of the number
of interfacial molecules times the average volume per molecule multiplied by an 
unknown factor whose value is between from 1/2 and 1.
This factor comes from considering two limiting cases. In the former case (1/2), the
centers of the interfacial molecules are assumed to be located at the cavity boundary
whereas in the latter one (1), the interfacial molecules are tangent to the 
cavity boundary.
If we assume an intermediate factor (3/4) the volume of the cavity calculated 
using the Voronoi tessellation is 0.98~nm$^3$, very close to the actual volume
of the artificial cavity.

\subsection{Metastable and unstable water}

\begin{center}
\begin{figure}[!ht]
\caption{Time evolution of the average volume per molecule for two runs of a 
point in the metastable region (T=280~K, p=-2250~bar). The sharp increase of
the volume at the end of the simulations correspond to cavitation events.}
\includegraphics*[clip,scale=0.8]{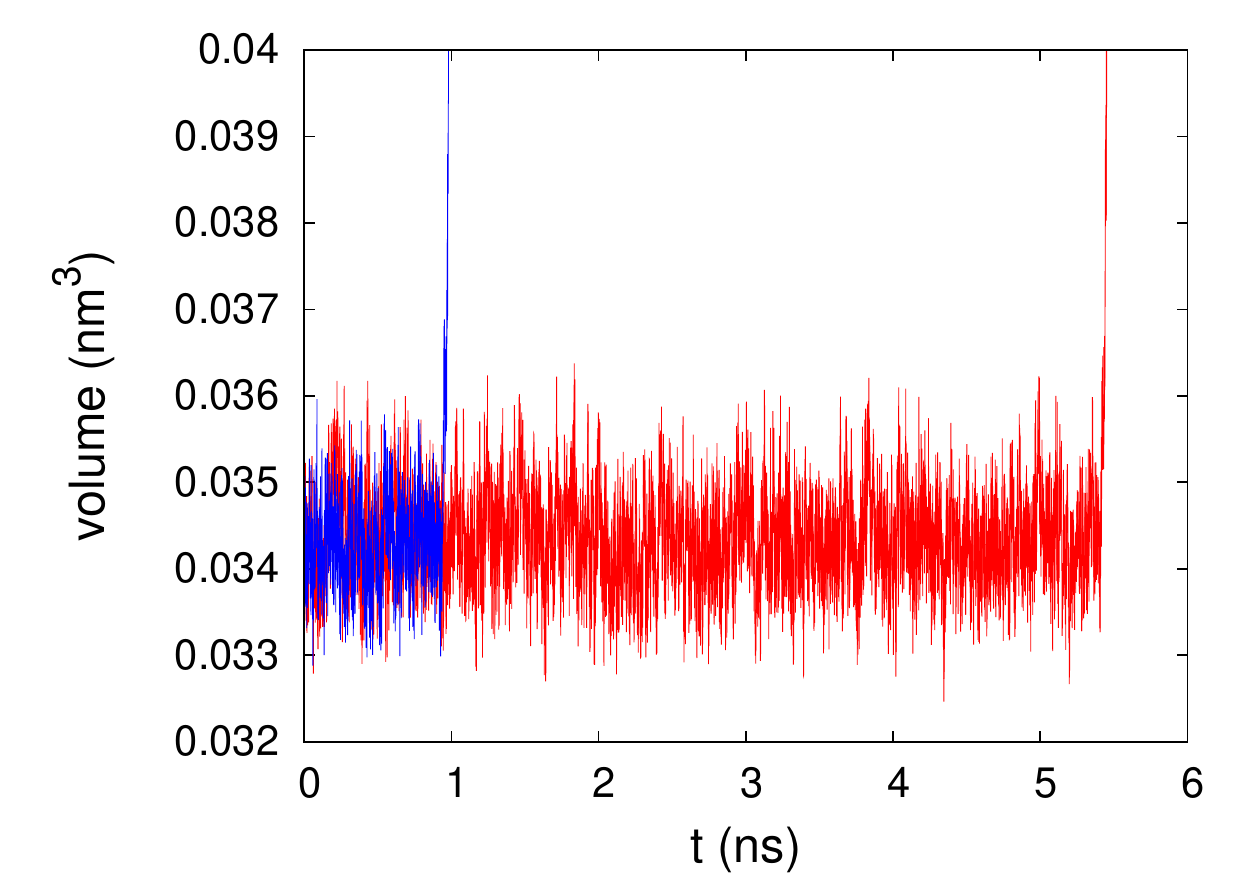}
\label{fig:volume-t}
\end{figure}
\end{center}

In this section we analyze the Voronoi tessellation of water in conditions far
from the thermodynamic stability.
This is performed at two state points at 280~K, one above (-2250~bar) and the other
below (-2630~bar) the liquid-vapor spinodal line corresponding to the
metastable and unstable regions, respectively.
In both cases the system cavitated spontaneously.
Figure~\ref{fig:volume-t} represents the average volume of the system for two
independent runs at T=280~K, p=-2250~bar.
Notice that the density fluctuations before cavitation takes place are very similar in both runs and that
the average volume is essentially constant along the simulations (before cavitation).
The resulting average volume per molecule for this state is $\bar{V}$=17.1nm$^3$.
Figure~\ref{fig:volume-t} also shows that cavitation happens 
stochastically: although the density fluctuations are akin in both runs, the
cavitation times are sensibly different.

Our main goal is to study bubble nucleation  from  metastable and unstable liquid water 
using a local order parameter (the volume of the largest bubble) whose construction is based 
on the Voronoi tessellation.
Therefore, as previously shown, we need to define the proper parameters of the Voronoi 
Polyhedra. This problem differs from that commented in the previous section (detection
of a cavity in a liquid) in several aspects.
First, the created cavity was static, {\em i.e.}, the molecules in the
liquid configurations had exactly the same positions as those in the system
with the cavity.
Thus, the VP properties of the molecules far away from the cavity were
exactly the same in both systems.
This allowed an unequivocal assignment of bulk and interfacial molecules and
enabled us to safely investigate the features of the Voronoi Polyhedra for the latter
particles. Bubbles forming in a metastable liquid are not static: they grow (and
disappear) until, eventually, the system cavitates.
A bubble may differ from an empty cavity because some reconstruction
must take place at the interface, and also because of the presence of a small
number of  vapor molecules inside the bubble (the ``cavity" is not empty
now).
However these  vapor molecules should have, by definition, large VP volumes so
the conditions imposed to tag a molecule as interfacial will also detect the
vapor ones.
Despite these differences, we may use the information obtained for the liquid
with a cavity in order to interpret the results for metastable water.

\begin{center}
\begin{figure}[ht]
\caption{Distribution of the reduced VP volumes for metastable water (T=280~K,
p=-2250~bar) compared to that of liquid water at ambient conditions with a
cavity}
\includegraphics*[clip,scale=0.8]{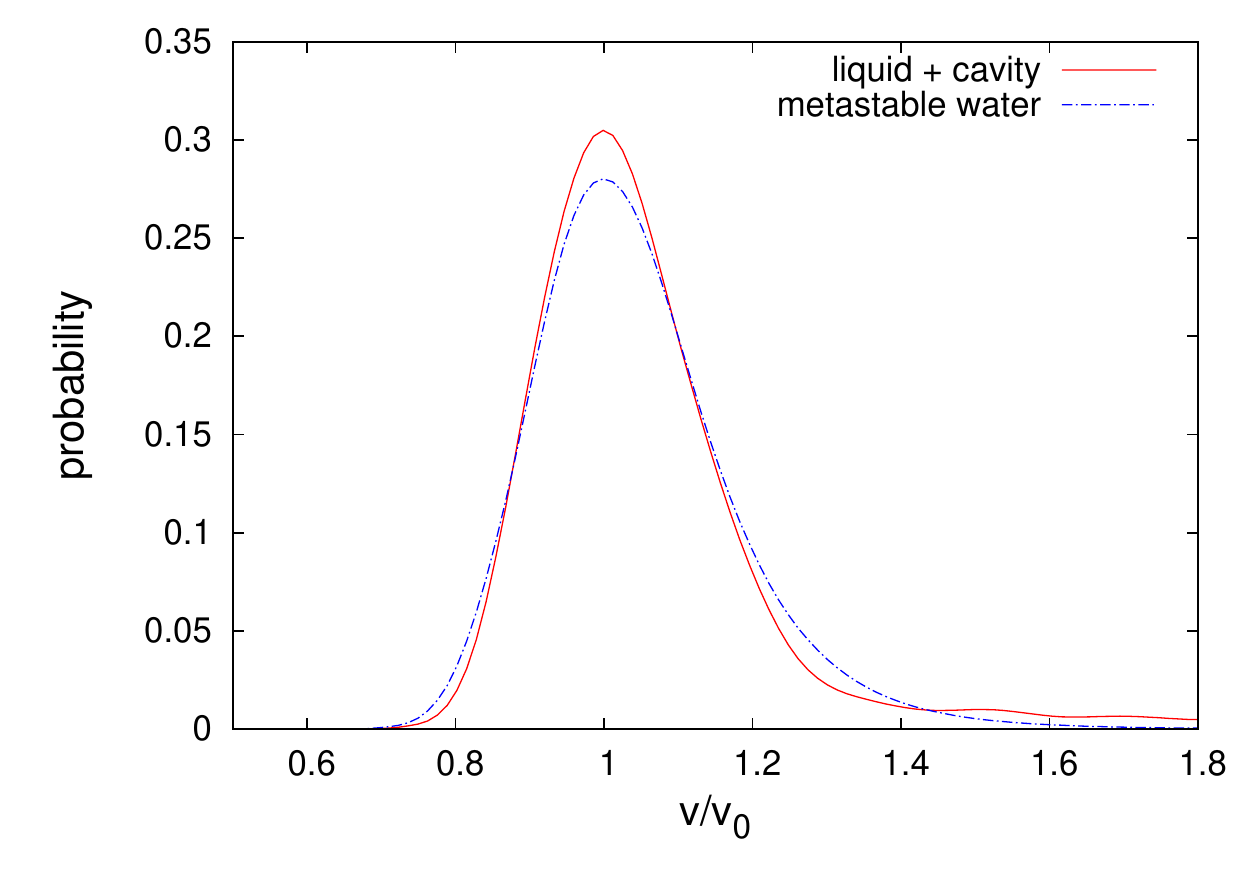}
\label{fig:vol_cavity-metastab}
\end{figure}
\end{center}

A closer look at the differences between the properties of the Voronoi 
tessellations of liquid water with a cavity and metastable water is shown
in Figure~\ref{fig:vol_cavity-metastab} where we present the distribution of
VP volumes.
Since the density is different, it seems important to scale the volumes using 
the average volume of the liquid molecules.
In practice, we have applied a factor so that the maximum of the distributions
appears at $V/V_0=1$.
The curves for both systems show minor differences.
Firstly, the distribution of volumes is wider for metastable water.
Besides, there is a minor but significant difference in the large volumes
region: the distribution in metastable water is smooth whereas the cavity
gives rise to the appearance of wiggles.
This seems to indicate that the abrupt inhomogeneity of the artificially
created cavity is not present in the bubbles whose interface is somewhat
blurred.
Figure~\ref{fig:vol_cavity-metastab} illustrates that the analysis made in the
previous section may also be acceptable for the detection of bubbles.
We only need to scale the parameters in order to account for the
differences in the average volume per molecule and the sharpness of the
distribution.

\begin{center}
\begin{figure}[!ht]
\caption{Anisotropic factor as a function of volume for metastable water
(red crosses) compared to that of liquid water (empty blue circles).
The black line $\alpha = 1.5-18*(V-0.048)$ separates liquid-like molecules
from vapor (interfacial) molecules.}
\includegraphics*[clip,scale=0.8]{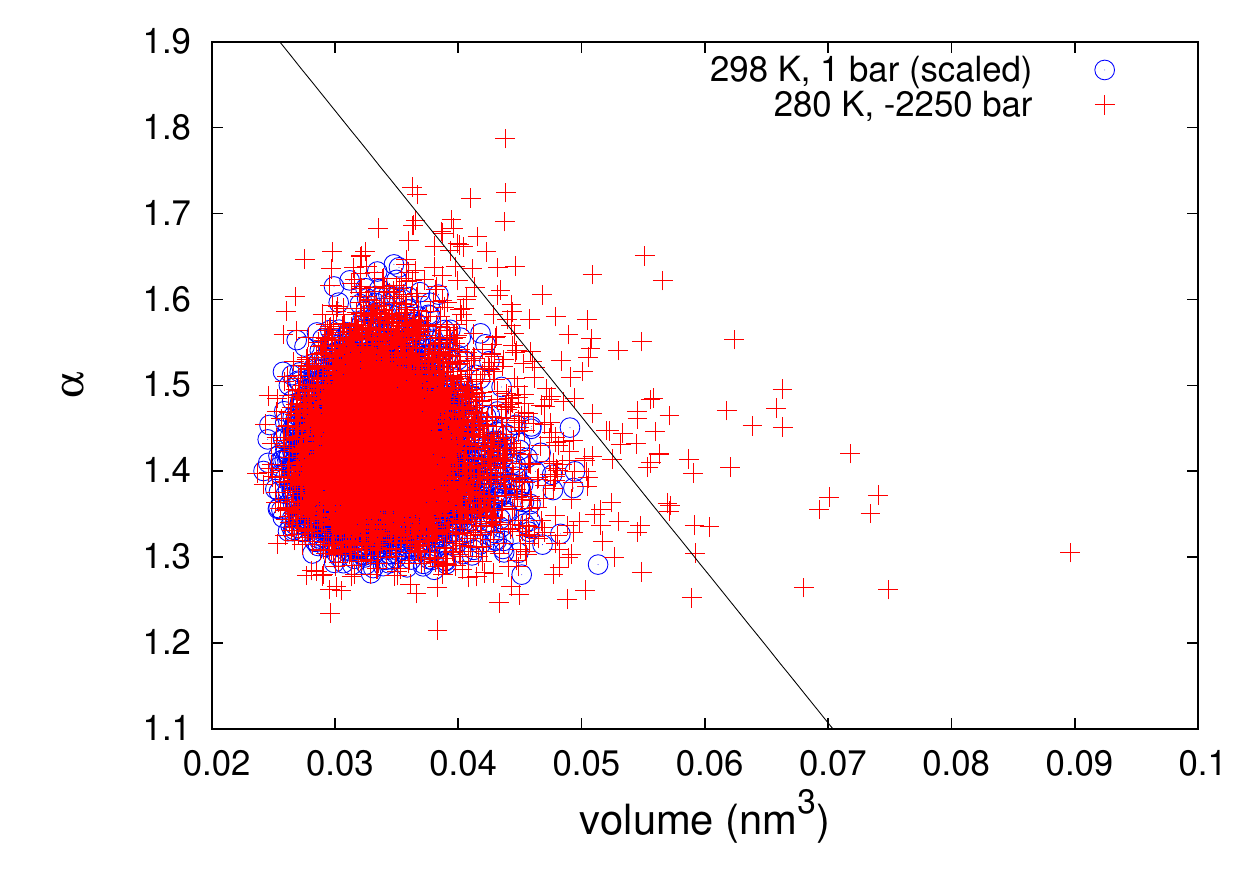}
\label{fig:vep_liq-metastab}
\end{figure}
\end{center}

In analogy to Fig.~\ref{fig:vep_liq-cavity}, Figure~\ref{fig:vep_liq-metastab}
shows the anisotropy of the Voronoi Polyhedra as a function of their volumes for 11 randomly
selected configurations of metastable water.
For comparison we have also plotted the data for liquid water
already shown in Fig.~\ref{fig:vep_liq-cavity} but scaled so that the mapping
of the distributions is maximized.
In accordance, we have also scaled the line separating the interfacial
molecules from the liquid molecules obtained for liquid water with a cavity.
The line contains the point (V=0.048~nm$^3$, $\alpha=1.5$) and its slope is
18~nm$^{-3}$.
It can be seen that the region to the left of the straight line for metastable
water is very similar to that for liquid water with scaled parameters.
This indicates that an analogous criterion (now the straight line $\alpha =
1.5-18*(V-0.048)$ shown in Fig.~\ref{fig:vep_liq-metastab}) may be used in both
systems to distinguish the liquid molecules from the interfacial (vapor) ones.
%

\begin{center}
\begin{figure}[!ht]
\caption{Anisotropic factor as a function of volume for unstable water
(T=280~K, p=-2630~bar) compared to that of liquid water with a cavity
(scaled)}
\includegraphics*[clip,scale=0.8]{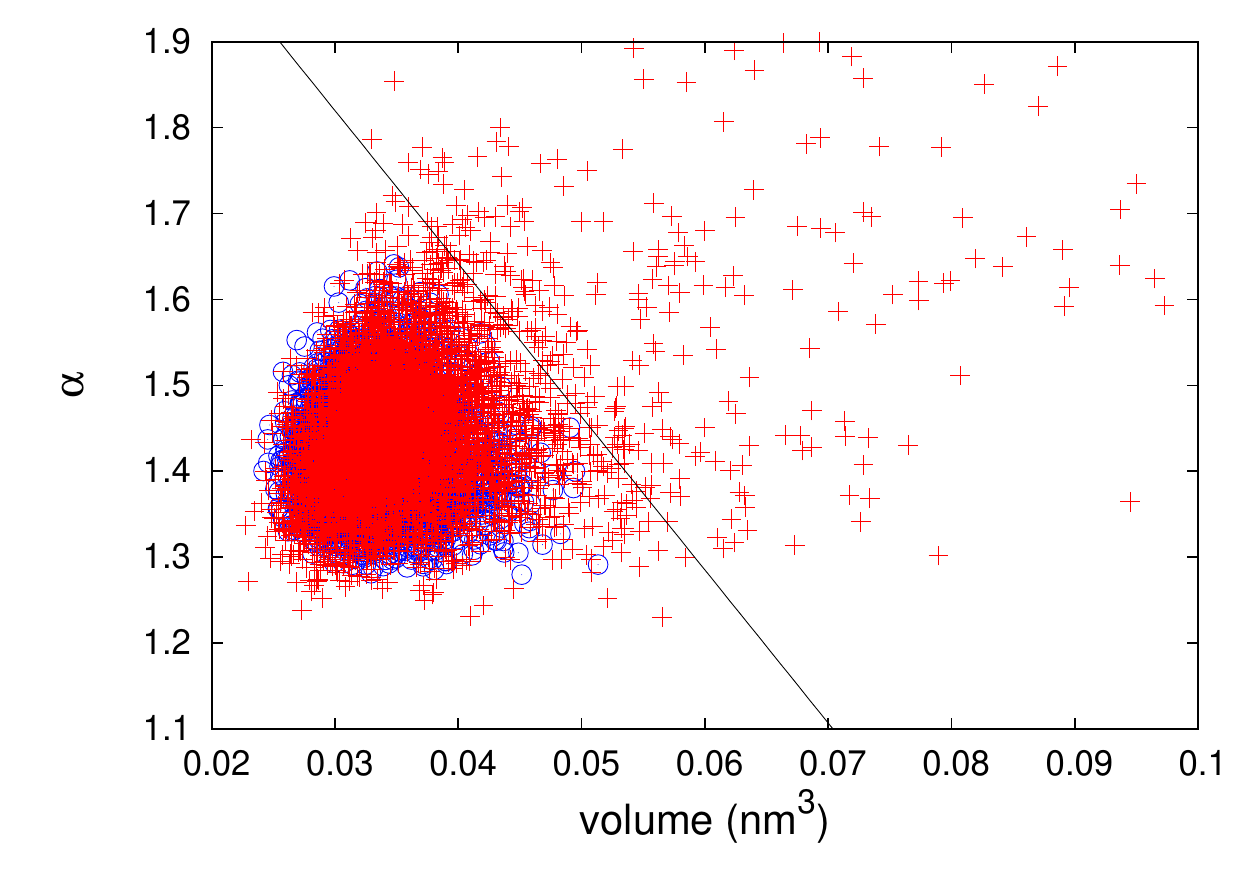}
\label{fig:vep_liq-unstab}
\end{figure}
\end{center}

The corresponding plot for water at a pressure (p=-2630~bar) below the spinodal
line (unstable water) is shown in Figure~\ref{fig:vep_liq-unstab}.
We have also included the data for liquid water using the same scaling as in
Figure~\ref{fig:vep_liq-metastab}.
The bulk region is again quite similar in both systems but the number of
interfacial molecules is much higher for unstable water.
Besides, in this case, the high VP volume of some molecules (up to 0.4~nm$^3$)
seem to indicate that the latter are actually vapor molecules (not shown in the
figure because, in order to compare with the previous ones, we use the same
axes scales in all of the $\alpha$-V plots).

\section{Mean first-passage times and nucleation rates}

\begin{center}
\begin{figure}[!ht]
\caption{Time evolution of the volume of the largest bubble for the runs of 
Fig.~\ref{fig:volume-t}} 
\includegraphics*[clip,scale=0.8]{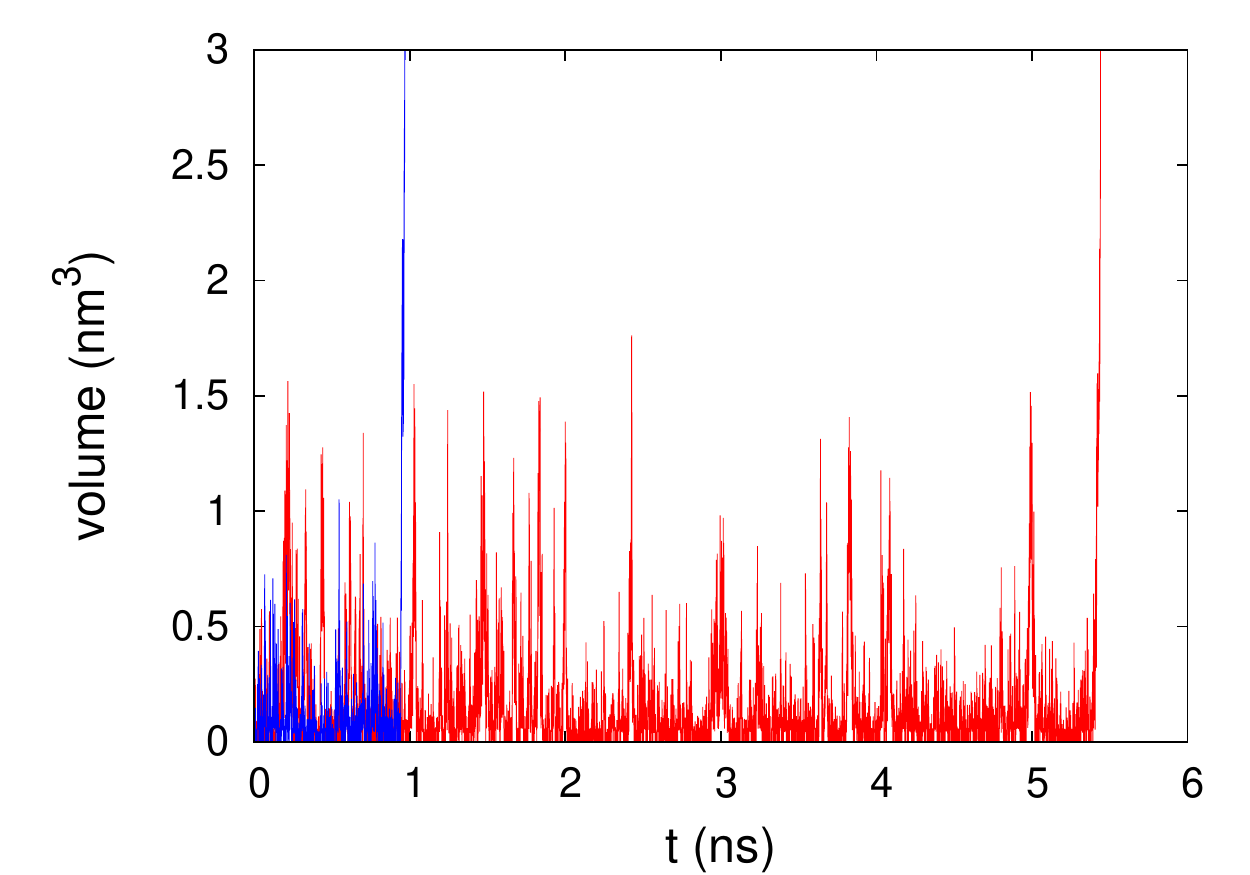}
\label{fig:lcl_metastab}
\end{figure}
\end{center}

Recently, Wedekind et al.\cite{wedekind07} proposed a formalism ---the
mean first-passage time (MFPT) method-- to investigate activated processes in
those cases where they can be observed spontaneously in molecular simulations.
In this work we have applied the method to investigate the nucleation rate of
bubble formation in metastable water using the volume of the largest bubble as the 
local order parameter.
Bubbles are obtained by clustering together the molecules tagged as  vapor
(or interfacial following the criteria established in the previous section)
that share a common face.
Their volumes are the sum of the VP volumes of the particles belonging
to a given cluster.
Only the largest bubble, {\em i. e.}, the cluster with the largest volume, is
required in the MFPT formalism.
For each trajectory, we evaluate the time it takes to reach a given volume for
the first time.
Being nucleation a stochastic event, the cavitation time varies significantly
from one run to another, as shown in Figure \ref{fig:lcl_metastab} for the system 
 at p=-2250 bar.  Moreover, Figure \ref{fig:lcl_metastab} also shows that at p=-2250 bar
the time required to form a critical bubble is much larger than the time  
the system takes to cavitate.
Once a bubble reaches a critical size it grows very quickly and destabilizes the
system that cavitates.
This indicates that nucleation and growth are not coupled.
A comparison of this plot with Figure~\ref{fig:volume-t} shows that the 
volume fluctuations are independent of the size of the largest bubble except 
when the bubble size exceeds the critical one.
In the latter case, the largest bubble grows very quickly and becomes
a significant part of the total volume (notice that the divergences in both
figures occur at the same time), and the system cavitates.

Fluctuations in Fig.~\ref{fig:lcl_metastab} already enable us to have a rough idea of the 
size of the critical cluster (around 1.5~nm$^3$). However, for a precise estimate
it is necessary to average the first passage times over a large number
of trajectories for a given thermodynamic state.
In this way we get a smooth MFPT curve from which it is possible to extract the 
rate and the size of the critical cluster.
In particular, the MFPT reported in this work is the outcome of 200 independent 
simulation runs.

\begin{center}
\begin{figure}[!ht]
\caption{A snapshot of the growth of a bubble. The image on the left
corresponds to a precritical bubble and the image on the right corresponds to a
postcritical one along the same trajectory (the configurations are separated
by 0.02ps).}
\includegraphics*[clip,scale=0.3]{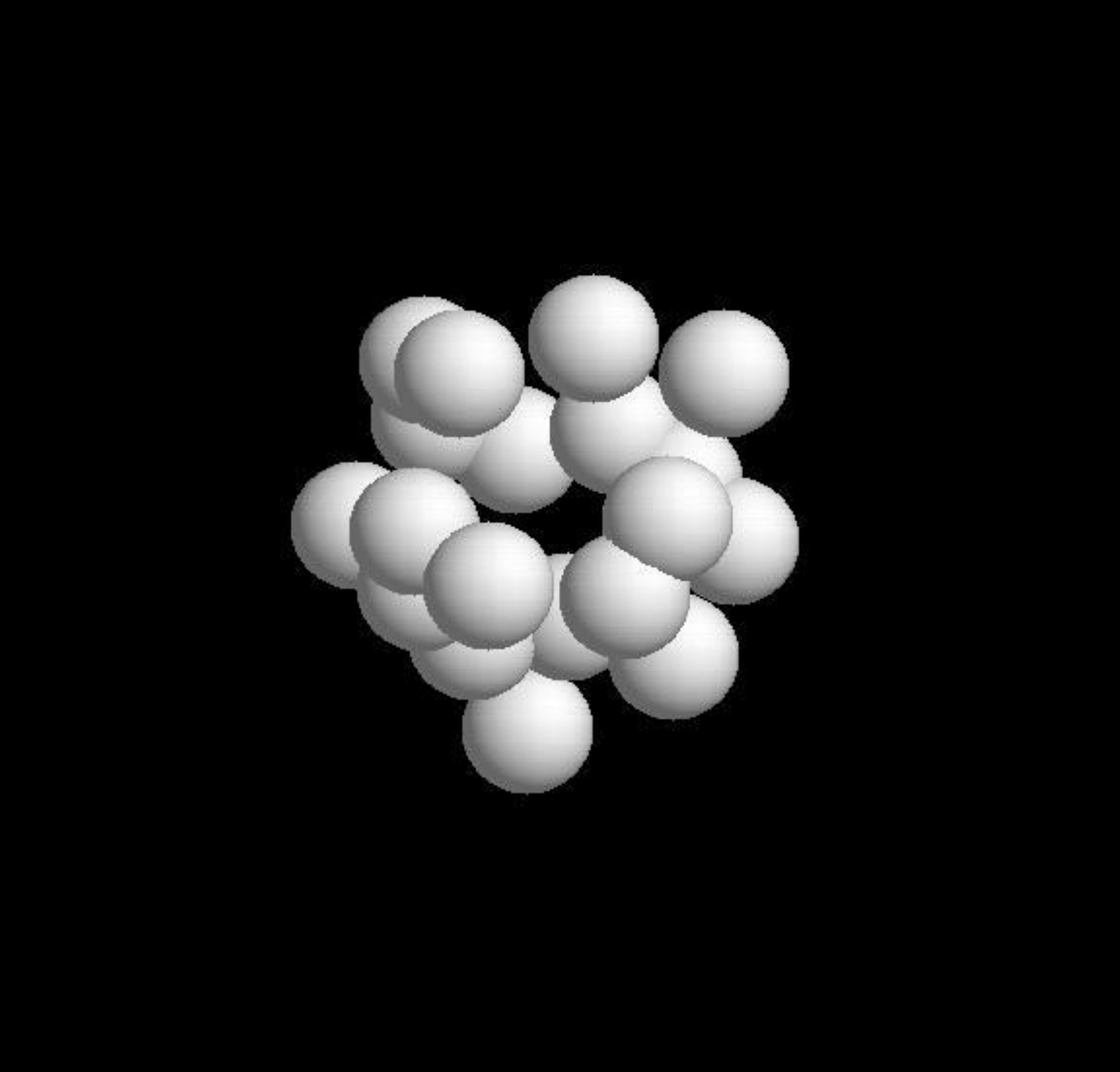}
\includegraphics*[clip,scale=0.3]{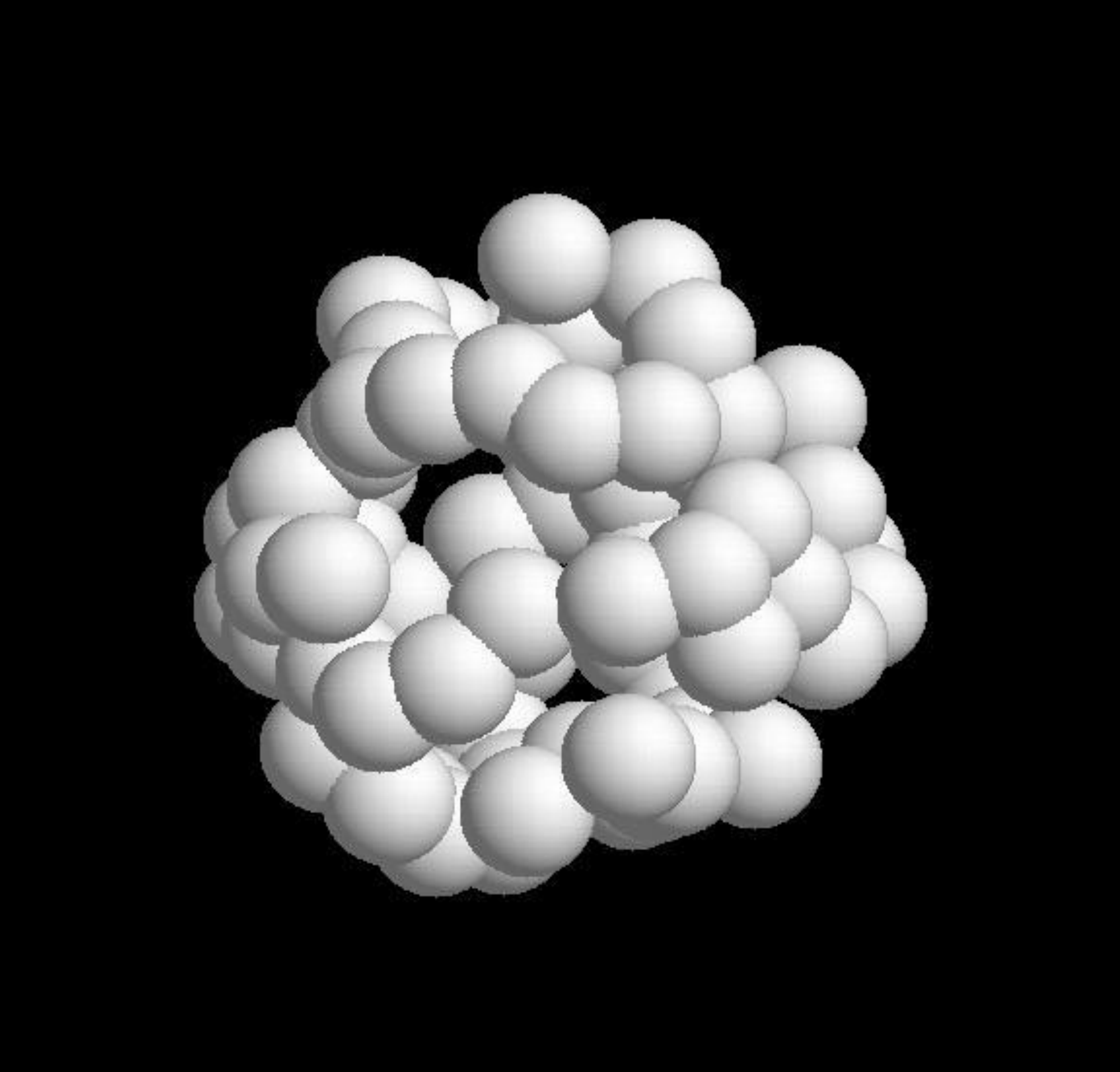}
\label{fig:twobubbles}
\end{figure}
\end{center}

\begin{center}
\begin{figure}[!ht]
\caption{Comparison of the MFPT obtained for metastable and unstable 
TIP4P/2005 water at 280 K.
Notice that a different scale is used for each of the curves (metastable liquid
on the left axis and unstable on the right axis.}
\includegraphics*[clip,scale=0.8]{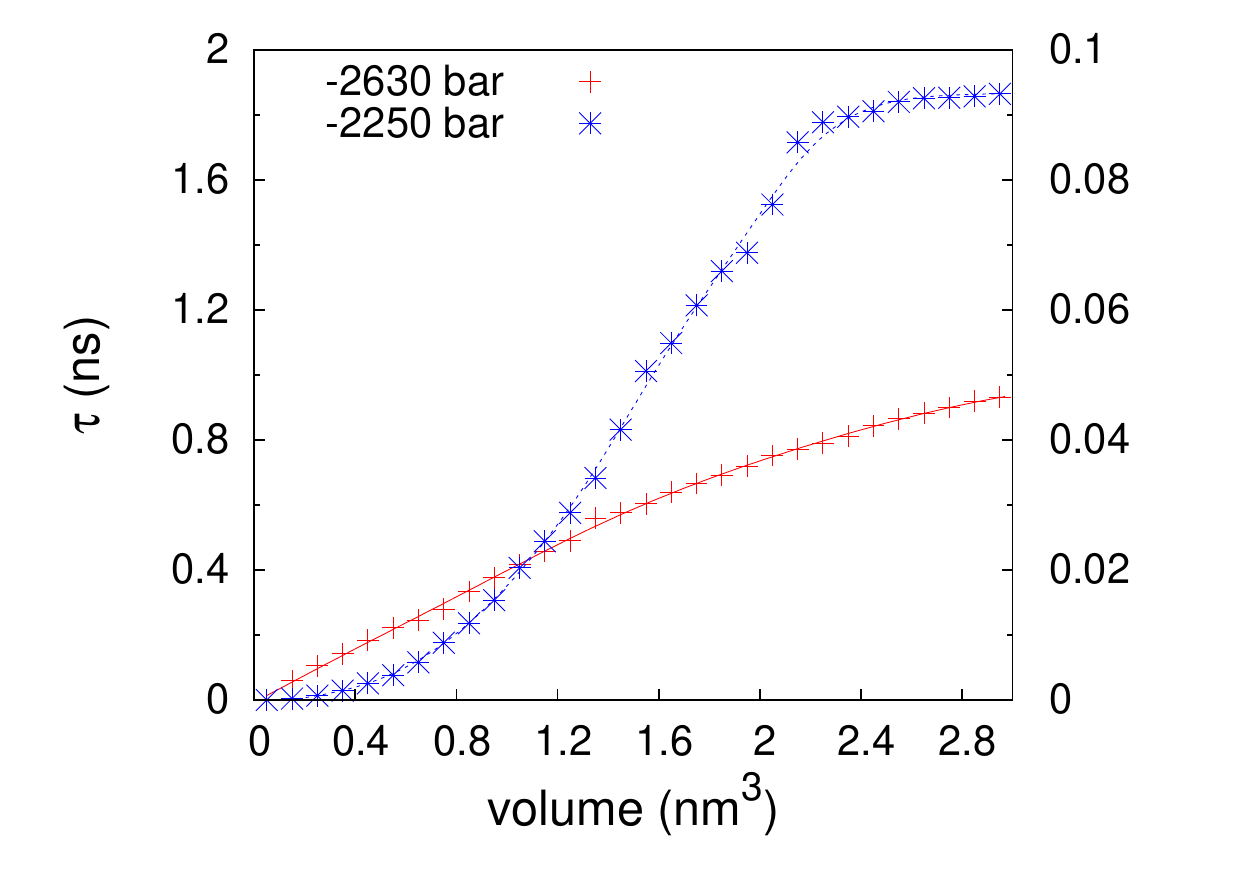}
\label{fig:mfpt_meta-unstab}
\end{figure}
\end{center}

Figure~\ref{fig:mfpt_meta-unstab} shows the comparison between the MFPT curve
for metastable water (above the spinodal) with that of unstable water (below
the spinodal).
Notice that a different time scale is used for each of the curves.
The nucleation behavior is quite different for these thermodynamic states.
MFPT for metastable water shows a typical sigmoidal shape which can be
fitted to a function of the form
\begin{equation}\label{MFPT1}
\tau(V)=(\tau_J/2)\{1+erf[c(V-V^*)]\}, 
\end{equation}
where $erf(x)$ is the error function.
Wedekind {\em et al.}\cite{wedekind07} have shown that the parameters of this
equation provide relevant physical information.
In particular, $V^*$ is the size of the critical cluster (location of the transition state) and $\tau_J$ is
the nucleation time which is related to the nucleation rate $J$ by
\begin{equation}
J=\frac{1}{\tau_J V_s},
\end{equation}
where $V_s$ is the system volume.
Applying Eq.~\ref{MFPT1} to the system at p=-2250 bar, we obtain 
$\tau_J=1.88$~ns, $V^*=1.52$~nm$^3$ and $c=1.26$.
From $\tau_J$ and $V^*$ it is immediate to evaluate the nucleation rate and the
radius of the critical cluster for which we obtain $J=0.031$~ns$^{-1}$nm$^{-3}$
and $r_{crit}=0.71$~nm (assuming the critical cluster to be spherical).
If we subtract the volume of the interfacial molecules, the volume of the critical
cluster becomes 0.82~nm$^3$, corresponding to a critical radius of 0.59~nm.
A visual inspection of the critical and postcritical bubbles show that they are
usually empty, i.e. no actual vapor molecules are within the bubble which is
essentially made only of interfacial molecules.
A snapshot of two bubbles, one slightly below the critical size and the other
one larger than the critical bubble, are shown if Fig~\ref{fig:twobubbles}.

Our results may be compared to those coming from the Classical Nucleation
Theory\cite{blander75} (CNT)
\begin{equation}
r_{crit}^{CNT}=\frac{2\gamma}{p_{sat}-p},
\end{equation}
\begin{equation}
J_{CNT}=\rho_N \left(\frac{2\gamma}{\pi m B}\right)^{1/2} e^{-\Delta G^{CNT}/kT},
\end{equation}
where $\gamma$ is the interfacial tension, $p_{sat}$ the coexistence pressure,
$m$ the mass of a molecule, $\rho_N$ the number density of the liquid and $B$
takes into account the mechanical equilibrium of the bubble (in cavitation 
experiments $B$=1), and 
$\Delta G^{CNT}=W(r_{crit}^{CNT})=(16\pi/3)(\gamma^3/(p_{sat}-p)^2$.
Assuming $\gamma \approx$ 73 mN/m\cite{vega07}, we obtain $r_{crit}^{CNT}\approx$ 
0.67~nm, and $J_{CNT}=1.7\times10^{-10}$ ~ns$^{-1}$nm$^{-3}$.
Thus, the prediction of the CNT for the critical radius is quite accurate
whereas the differences in the nucleation rates are about eight orders of
magnitude.
This is in agreement with the observations made for the Lennard-Jones 
system.\cite{zeng91}.

On the contrary, the MFPT curve for unstable water is a monotonous function
without an inflection point nor a plateau indicating that, in this case,
nucleation and growth are coupled.
A number of bubbles are quickly formed. When these bubbles grow, some of 
them eventually merge together giving rise to larger bubbles and the 
growth-coalescence process continues until cavitation (spinodal decomposition).
This means that there is no nucleation free-energy barrier which is
consistent with the fact that the system's pressure is below the spinodal.

\begin{center}
\begin{figure}[!ht]
\caption{MFPT for metastable water using two different VP parameters to
distinguish liquid from `` vapor" molecules.}
\includegraphics*[clip,scale=0.8]{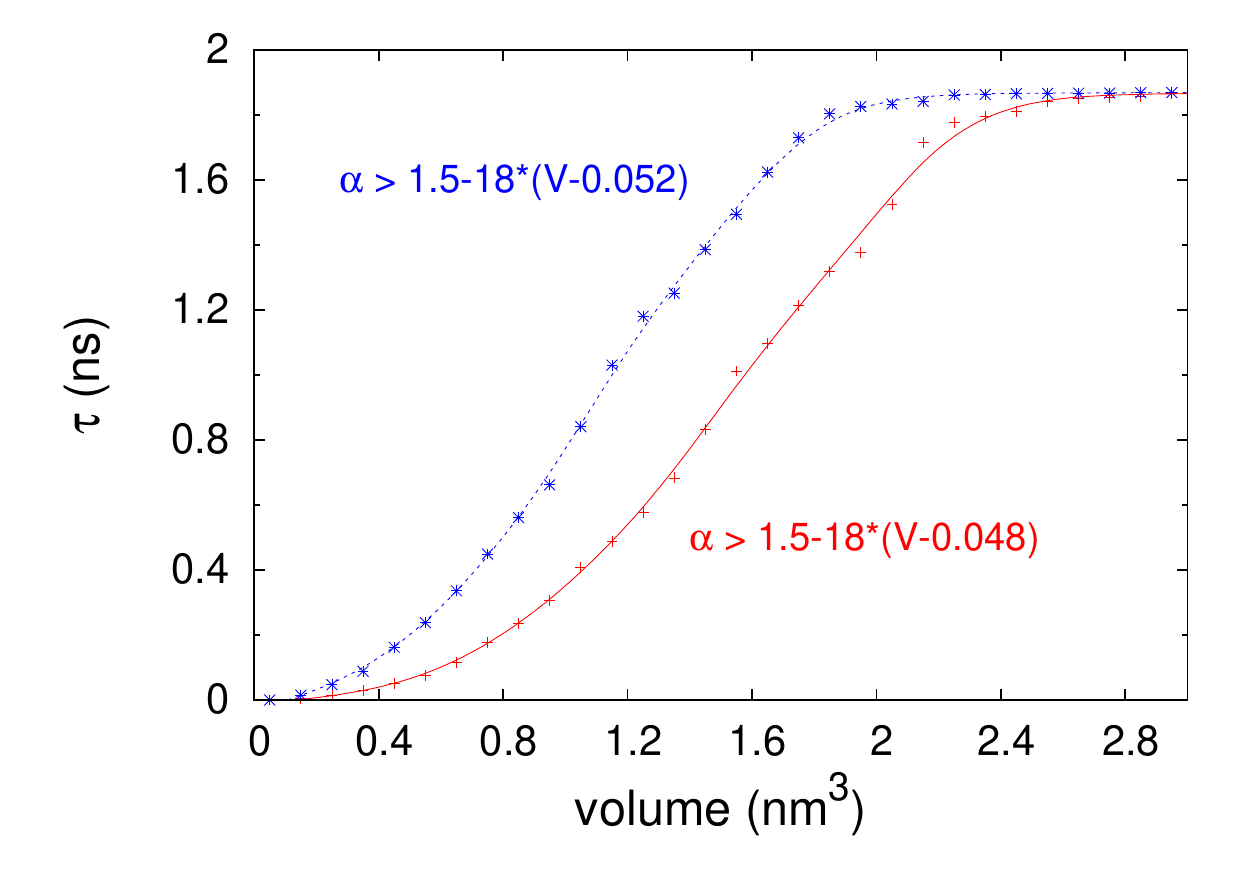}
\label{fig:mfpt_metastab_v048-052}
\end{figure}
\end{center}

\begin{center}
\begin{figure}[!ht]
\caption{MFPT for metastable water for two different sampling times
(configurations are analyzed every $\Delta$t~ps).}
\includegraphics*[clip,scale=0.8]{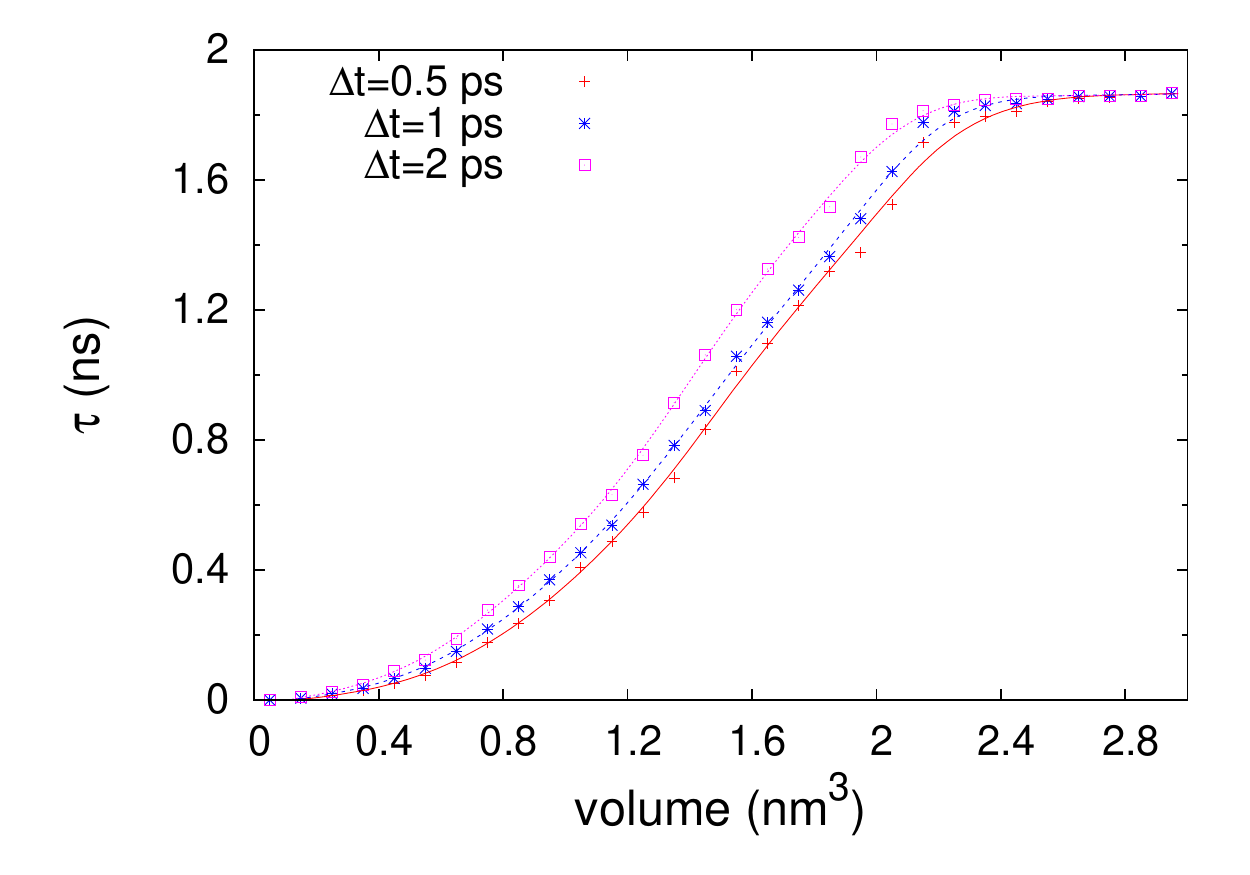}
\label{fig:mfpt_metastab_ev}
\end{figure}
\end{center}

We now study how the choice of parameters affect the final results.
Figure~\ref{fig:mfpt_metastab_v048-052} shows the MFPT using two different
criteria to distinguish liquid from vapor molecules. 
In both cases, we tag a molecule as  vapor if its VP anisotropy, $\alpha$, 
is above a straight line defined in terms of a pivotal point and a slope in the
$\alpha$-V plane.
One of the curves correspond to the parameters provided by the study made in 
the previous section, $\alpha > 1.5-18*(V-0.048)$, and the other one
has been obtained with $\alpha > 1.5-18*(V-0.052)$, a quite restrictive 
condition (see Fig.~\ref{fig:vep_liq-metastab}).
A more restricted choice implies a smaller number of molecules fulfilling
the condition and, as a result, the MFPT curve is shifted toward lower volumes.
The size of the critical bubble is then slightly dependent on the choice of
parameters.
Interestingly, the value of $\tau_J$ (and, thus, the nucleation rate) does not
depend at all on the definition of the order parameter.
This is clear a demonstration of the robustness of the MFPT technique.
Since the trajectories changes very little from one configuration to the next
one, we only analyze them every $\Delta t$~ps.
Figure~\ref{fig:mfpt_metastab_ev} shows the effect of modifying the sampling
time  $\Delta t$.
Notice that sampling times sensibly larger than the simulation timestep 
can be used.
In fact, an increase by a factor of four in $\Delta t$ has a little effect
on the size of the critical cluster and a null influence on the nucleation
rate.

\section{Discussion and conclusions}

In this work we set up a Voronoi-based tessellation procedure 
to determine the volume of the largest vapor bubble that allows us to study 
bubble nucleation in an over-stretched TIP4P/2005 metastable water.
The parameters of the Voronoi polyhedra that allow to distinguish bulk (liquid)
particles from interfacial (or  vapor) ones have been obtained from the
investigation of liquid configurations in which a cavity has been artificially created.
It has been shown that best choice of parameters involves a combination of 
the Voronoi Polyhedra volume and  nonsphericity factor $\alpha$.
The study allows to reduce to a minimum the intrinsic arbitrariness in fixing
the parameters required to detect the growth of bubbles in the metastable
liquid.  As a consequence the nucleation rate is completely independent on the
parameters chosen to define the local order and the critical cluster size is
only moderately dependent on them (the MFPT technique also contributes to this
achievement).

One important open question is the effect of the finite-size. It has been pointed
that the MFPT analysis is less sensitive to finite size effects than other
approaches because it only requires a box large enough to contain several times
the critical cluster.\cite{perez11}
Since our calculated volume of the critical cluster for metastable TIP4P/2005
water is 1.5 nm$^3$ and the average system volume is about 17 nm$^3$, we conclude 
that  our chosen system fulfils the required condition and we can neglect 
(at least, at this stage) finite-size effects.

As in simple fluids, the bubble nucleation rate  estimated with Classical
Nucleation Theory turns out to be much lower (8 orders of magnitude) than the
one computed using MFPT. This could be due to the fact that CNT neglects
curvature corrections to the surface free-energy of the bubble.  
When analyzing the mechanism of bubble formation above and below the spinodal, 
we observe clear differences. 
In the simulations of the liquid above the spinodal line, 
one bubble grows larger than all others in a fairly compact 
shape whereas, for the liquid below the spinodal line, we detect the
formation of  few bubbles which, eventually, merge to form larger ones 
(spinodal decomposition).
However, above the spinodal we have also detected bubbles with different
topologies, proving that small bubbles are strongly fluctuating objects. In the
coming future, we plan to attempt a systematic characterization of the topology
of the bubbles, as we believe it might play an important role in the cavitation
process.

In general, the Voronoi tessellation  is a relatively CPU-time consuming method.
For a deeper investigation of the bubble nucleation mechanism it seems interesting to
devise alternative ways of partitioning and calculating the bubble volume.
Therefore, the results presented in this work can be seen as a benchmark for
further bubble nucleation studies.

\section*{Acknowledgments}
This work has been funded by grants FIS2010/16159 of the MEC, P2009/ESP-1691
of CAM and the Marie Curie Integration Grant PCIG-GA-2011-303941 (ANISOKINEQ) 
C.V. also acknowledges financial support from a Juan de La Cierva Fellowship. 
We acknowledge Carlos Vega for helpful discussions and a critical reading of
the manuscript.

\bibliographystyle{apsrev}

\end{document}